\documentstyle[stwol,psfig]{article}

\def\e{\epsilon}

\def\e3{$\epsilon_3$}

\def\ch2{$\chi^2$}

\def\co#1{{\ifmmode{\cal O}_{#1}\else${\cal O}_{#1}$\fi}}

\newdimen\unit
\def\point#1 #2 #3{\vbox to0pt{\kern-#2\unit
  \hbox{\kern#1\unit#3}\vss}
 \nointerlineskip}

\newcommand{\be}{\begin{equation}}
\newcommand{\ee}{\end{equation}}
\newcommand{\bea}{\begin{eqnarray}}
\newcommand{\eea}{\end{eqnarray}}

\begin{document}
\begin{titlepage}
 \hfill       OHSTPY-HEP-T-98-030\\
\mbox{ } \hfill           NUHEP-TH-99-73\\
  \mbox{ } \hfill      March 1999 \\

\vspace{2.0cm}

\begin{center}

\renewcommand{\thefootnote}{\fnsymbol{footnote}}

  {\Large\bf
      Neutrino Oscillations in a Predictive SUSY GUT}

  \vspace{1cm}

  {\Large\bf
    T. Bla\v{z}ek$^\dagger$$^\ddagger$,  S. Raby$^*$ and K. Tobe$^*$}

    \bigskip
{\em $^\dagger$Department of Physics and Astronomy,
Northwestern University,
Evanston, IL 60208

$^*$Department of Physics,
The Ohio State University,
174 W. 18th Ave.,
Columbus, Ohio  43210}

  \vspace{1cm}
{\bf Abstract}
\end{center}

In this letter we present a predictive SO(10) SUSY GUT with flavor symmetry
U(2)$\times$U(1) which has several nice features.  We are able to fit fermion
masses and mixing angles, including recent neutrino data, with 9 parameters
in the
charged fermion sector and 4 in the neutrino sector.  The flavor symmetry
plays a
preeminent role --

(i) The model is ``natural" -- we include all terms allowed by the
symmetry.  It restricts the number of arbitrary parameters and enforces
many zeros
in the effective mass matrices.

(ii)  Flavor symmetry breaking from U(2)$\times$U(1) $\rightarrow$ U(1)
$\rightarrow$ nothing generates the family hierarchy.  It also constrains
squark
and slepton mass matrices, thus ameliorating flavor violation resulting from
squark and slepton loop contributions.

(iii) Finally, it naturally gives large angle $\nu_\mu - \nu_\tau$ mixing
describing atmospheric neutrino oscillation data and small angle $\nu_e -
\nu_{s}$
mixing consistent with the small mixing angle MSW solution to solar
neutrino data.

\vfill
$^\ddagger$ {\em
           On leave of absence from
           Faculty of Mathematics and Physics,
           Comenius Univ., Bratislava, Slovakia}
\setcounter{footnote}{0}
\renewcommand{\thefootnote}{\arabic{footnote}}

\end{titlepage}
%
%

\title{Neutrino Oscillations in a Predictive SUSY GUT}

\author{T. Bla\v{z}ek$^\dagger$,  S. Raby$^*$ and K. Tobe$^*$}

\address{$^\dagger$Department of Physics and Astronomy, Northwestern
University,
Evanston, IL 60208 \\
                   and Faculty of Mathematics and Physics,
           Comenius Univ., Bratislava, Slovakia\\
$^*$Department of Physics, The Ohio State University,
174 W. 18th Ave., Columbus, Ohio  43210}


\twocolumn[\maketitle\abstracts{
In this letter we present a predictive SO(10) SUSY GUT with flavor symmetry
U(2)$\times$U(1) which has several nice features.  We are able to fit fermion
masses and mixing angles, including recent neutrino data, with 9 parameters in
the charged fermion sector and 4 in the neutrino sector.  The flavor symmetry
plays a preeminent role --
(i) The model is ``natural" -- we include all terms allowed by the
symmetry.  It restricts the number of arbitrary parameters and enforces many
zeros in the effective mass matrices.
(ii)  Flavor symmetry breaking from U(2)$\times$U(1) $\rightarrow$ U(1)
$\rightarrow$ nothing generates the family hierarchy.  It also constrains
squark
and slepton mass matrices, thus ameliorating flavor violation resulting from
squark and slepton loop contributions.
(iii) Finally, it naturally gives large angle $\nu_\mu - \nu_\tau$ mixing
describing atmospheric neutrino oscillation data and small angle $\nu_e -
\nu_{s}$
mixing consistent with the small mixing angle MSW solution to solar neutrino
data.}]

\section{Introduction}

Solar~\cite{solar}, atmospheric~\cite{atmos} and accelerator~\cite{LSND}
neutrino data strongly suggest that neutrinos have small masses and
non-vanishing mixing angles.  This hypothesis is also constrained
by reactor~\cite{chooz} based experiments. In the near future, many more
experiments will test the hypothesis of neutrino masses.  In addition, a
neutrino mass necessarily implies new physics beyond the Standard Model.
Thus there is great excitement, both experimental and theoretical, in this
field.

Phenomenological neutrino mass models~\cite{massmatrices} are  designed to
reproduce the best fits to all or some of the neutrino data.   These
models are only constrained by how much of the neutrino data one wants to fit.
Three neutrino models with 3 active neutrinos  ($\nu_e,\;
\nu_\mu \;\; \rm and \;\; \nu_\tau$) are consistent with solar~\cite{solar}
and atmospheric~\cite{atmos} neutrino oscillations, while four neutrino
models, including a sterile (or electroweak singlet) neutrino ($\nu_s$), are
consistent with
solar, atmospheric and LSND~\cite{LSND} neutrino experiments.  There are also
6 neutrino models, with 3  active and 3 sterile
neutrinos, motivated by complete family symmetry~\cite{geisser}.

It is important to address the theoretical question; to what extent can
this new data on neutrino masses and mixing angles constrain the physics
beyond the Standard Model; in particular, theories of fermion masses.  Since
any number of sterile neutrinos may mix with the 3 active neutrinos, even
in a grand unified theory, it may always be possible to fit neutrino data
without ever constraining the charged fermion sector of the theory.
This would be an unfortunate circumstance.  It is the purpose of this paper,
however, to show that in any ``predictive" theory of charged fermion masses,
the neutrino sector is severely constrained.

By a ``predictive" model of fermion masses we mean --
\begin{itemize}
\item  the Lagrangian is ``natural" containing all terms consistent
with the symmetries  and particle content of the theory.
\item In addition there are necessarily grand unified [GUT] gauge
symmetries as well as family symmetries which restrict the form of the Yukawa
matrices~\cite{familysymmetry,u2symmetry,so10u2}; thereby greatly reducing the
number of arbitrary parameters.
\item  In supersymmetric [SUSY] theories, these same family symmetries can
usefully constrain the form of soft SUSY breaking squark and slepton masses as
well~\cite{familysymmetry,u2symmetry,so10u2}; thus ameliorating the problem of
large flavor violation in SUSY~\cite{flavorviolation}.
\end{itemize}

{\em In this letter, we demonstrate that these same family symmetries greatly
restrict the form of neutrino masses and mixing.  Hence neutrino data can
greatly constrain any predictive theory of fermion masses.}

We show this in the context of a particular SO(10)  SUSY
GUT which fits charged fermion masses and mixing angles
well.  SUSY GUTs are very attractive.  They successfully predict the
unification of gauge couplings observed at LEP~\cite{drw,recentsusyguts}.
In SO(10) one family $\{ q,\ \bar u,\ \bar d,\ l,\ \bar e,\ \bar \nu \}$ fits
into the {\bf 16} dimensional spinor representation of the
group~\cite{georgi}.
Thus up, down, charged lepton and Dirac neutrino mass matrices are related.

Of course, the last comment leads to the generic problem for any GUT
description
of neutrino masses.  Atmospheric neutrino data~\cite{atmos} requires large
mixing between $\nu_{\mu}$ and $\nu_x$, where $\nu_x$ is any neutrino
species,  other than $\nu_e$~\cite{atmos,chooz}.  Solar neutrino data as well
can have a  large mixing angle solution.  Thus lepton mass matrices must give
large mixing  angles in sharp contrast to quark mass matrices which give
small CKM mixing angles.

We consider an SO(10)$\times$U(2)$\times$U(1) model of fermion masses.
This theory is a modification of the SO(10)$\times$U(2) model of Barbieri,
Hall, Raby and Romanino [BHRR]~\cite{so10u2}.  The modifications only affect
the results for neutrinos. Alternate descriptions of neutrinos in the context
of U(2) family symmetry can be found in recent articles~\cite{u2neutrino}.
In section 2, we give the superspace potential and the resulting quark and
lepton Yukawa matrices.  We then give the results for charged fermion masses
and mixing angles.  In section 3, we describe the neutrino sector;  giving
our fits for solar and atmospheric neutrino oscillations and predictions
for future
experiments.  We are not able to accomodate LSND. Our conclusions are in
section 4.

\section{An SO(10)$\times$U(2)$\times$U(1) model}

The three families of fermions are contained in $16_a, \, a = 1,2;$ and
$16_3$ where
$a$ is a U(2) flavor index.  [Note U(2) = SU(2) $\times$ U(1)$'$ where the
U(1)$'$ charge is +1 ($-1$) for each upper (lower) SU(2) index.]  At tree
level, the third family of fermions couples to a $10$ of Higgs with coupling
$\lambda \; 16_3 \; 10 \; 16_3$ in the superspace  potential.  The Higgs and
$16_3$ have zero charge under both U(1)s, while $16_a$ has charge $-1$ and thus
does not couple to the Higgs at tree level.~\footnote{There are in fact three
additional U(1)s implicit in the superspace potential (eqn. \ref{eq:W}).
These are a Peccei-Quinn symmetry in which all 16s  have charge +1,  all
$\overline{16}$s have charge $-1$, and 10 has charge $-2$; a flavon symmetry in
which ($\phi^a, \; S^{a\, b}, \; A^{a \, b}$) and $M$ have charge +1 and
$\bar \chi_b$ has charge $-1$; and an R symmetry in which all chiral
superfields have charge +1. The flavor symmetries of the theory may be
realized as either global or local symmetries.  For the purposes of this
letter, it is not necessary to specify which one.
However, if it is realized locally, as might be  expected from string
theory, then
not all of the U(1)s are anomally free.   We would then need to specify the
complete
set of anomally free U(1)s.}

Three superfields ($\phi^a, \; S^{a\, b} = S^{b\, a}, \; A^{a \, b} =
-A^{b\, a}$) are
introduced to spontaneously break U(2)$\times$U(1) and to generate Yukawa
terms giving mass to the first and second generations.  The fields
($\phi^a, \;
S^{a\, b}, \; A^{a \, b}$) are SO(10) singlets with U(1) charges \{0, 1, 2\},
respectively.  The vacuum expectation values [vevs] ($\phi^2  \sim
S^{2\, 2}  \sim  \epsilon M_0^2/\langle 45 \rangle$) break U(2)$\times$U(1) to
$\tilde U(1)$ and  ($A^{1 \, 2} \sim \epsilon'
M_0$) completely.  In
this model, second generation masses are of order $\epsilon$, while first
generation masses are of order $\epsilon'^2/\epsilon$.

The superspace potential for the charged fermion sector of this theory,
including
the heavy Froggatt-Nielsen states~\cite{fn}, is given by
\begin{eqnarray}  W \supset & 16_3 \; 10\; 16_3 \;\; +\; \;  16_a \; 10 \;
\chi^a
& \label{eq:W} \\
& + \;\; \bar \chi_a \; (M \; \chi^a \; +\; \phi^a \; \chi \; +\; S^{a\; b} \;
 \chi_b \; + \;  A^{a\; b} \; 16_b) & \nonumber \\
&+ \;\; \bar \chi^a \; (M' \; \chi_a \;\; + \;\; 45 \; 16_a) & \nonumber \\
& + \;\; \bar \chi \; (M'' \; \chi \;\; + \;\; 45 \; 16_3) & \nonumber
\end{eqnarray}
where $M = M_0 (1\;\;+ \;\;  \alpha \; X \;\; +\;\; \beta \; Y)$.
$X,\; Y$ are SO(10) breaking vevs in the adjoint representation with $X$
corresponding to the  U(1) in SO(10) which preserves SU(5),  $Y$ is
standard weak
hypercharge and   $\alpha, \; \beta$ are arbitrary parameters.  The field $45$
is assumed to obtain a vev in the B - L direction.  Note, this theory differs
from BHRR~\cite{so10u2} in that the fields $\phi^a$ and $S^{a\, b}$ are now
SO(10) singlets (rather than SO(10) adjoints) and the SO(10) adjoint quantum
numbers of these fields, necessary for acceptable masses and mixing angles,
has been made explicit in the field $45$ with U(1) charge
1.~\footnote{This change
(see BHRR~\cite{so10u2}) is the reason for the  additional U(1).}  This
theory thus
requires much fewer SO(10) adjoints.   Moreover our neutrino mass solution
depends
heavily on this change.

The effective mass parameters $M_0, \; M', \; M''$ are SO(10) invariants.
The scales are assumed to satisfy  $M_0 \sim M' \sim  M'' \gg
\langle \phi^2 \rangle \sim \langle S^{2\, 2} \rangle \gg \langle A^{1\, 2}
\rangle $ where $M_0$ may be of order the GUT scale.
In the effective theory below $M_0$, the Froggatt-Nielsen states  \{$\chi,\,
\bar \chi,\ \chi^a,\, \bar \chi_a,\  \chi_a,\, \bar \chi^a$\}
may be integrated out, resulting in the effective Yukawa matrices for up
quarks,
down quarks, charged leptons and the Dirac neutrino Yukawa matrix given by
(see fig. 1)
\begin{eqnarray}
Y_u =&  \left(\begin{array}{ccc}  0 & \epsilon' \rho & 0 \\
                          - \epsilon' \rho &  \epsilon \rho & r \epsilon
T_{\bar
u}     \\
                      0  & r \epsilon T_Q& 1 \end{array} \right) \; \lambda &
\nonumber \\
Y_d =&  \left(\begin{array}{ccc}  0 & \epsilon'  & 0 \\
                          - \epsilon'  &  \epsilon  &  r \sigma \epsilon
T_{\bar
d}\\
                      0  &  r \epsilon T_Q & 1 \end{array} \right) \; \xi
& \label{eq:yukawa}
  \\
Y_e =&  \left(\begin{array}{ccc}  0 & - \epsilon'  & 0 \\
                           \epsilon'  &  3 \epsilon  &  r \epsilon T_{\bar
e} \\
                      0  &  r \sigma \epsilon T_L & 1 \end{array} \right)
\; \xi &
 \nonumber \\
Y_{\nu} =&  \left(\begin{array}{ccc}  0 & - \omega \epsilon'  & 0 \\
                  \omega \epsilon'  &  3 \omega \epsilon  & {1 \over 2}
\omega r
\epsilon T_{\bar \nu}
\\
                      0  &  r \sigma \epsilon T_L& 1 \end{array} \right) \;
\lambda &
 \nonumber
\end{eqnarray}
with  \begin{eqnarray} \omega = {2 \, \sigma \over 2 \, \sigma - 1}
\label{eq:omega} \end{eqnarray} and
\begin{eqnarray} T_f  = & (\rm Baryon\# - Lepton \#) &
\label{eq:Tf} \\
\rm for & f = \{Q,\bar u,\bar d, L,\bar e, \bar \nu\}.& \nonumber
\end{eqnarray}

\begin{figure}
\vspace{-1cm}
	\centerline{ \psfig{file=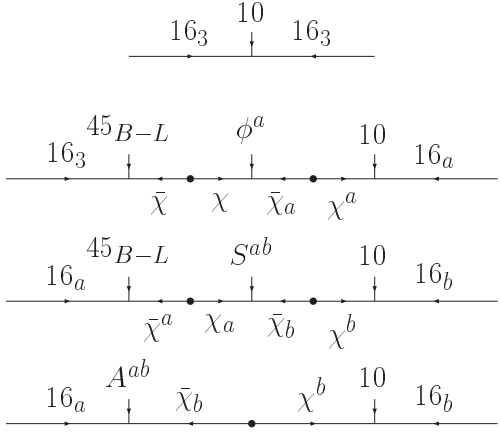,width=10cm,rheight=7cm}}
\caption{Diagrams generating the Yukawa matrices}
\end{figure}

In our notation, fermion doublets are on the left and singlets are on the
right.  Note,
we have assumed that the Higgs doublets of the minimal supersymmetric standard
model[MSSM] are contained in the
$10$ such that
$\lambda \;10 \supset \lambda \; H_u \; + \; \xi \; H_d$.  We can then
consider two
important limits ---  case (1)  $\lambda = \xi$ (no Higgs mixing) with large
$\tan\beta$, and case (2)  $\lambda \gg \xi$ or small $\tan\beta$.

\subsection{Results for Charged Fermion Masses and Mixing Angles}

\protect
\begin{table}
\caption[8]{
{\bf Charged fermion masses and mixing angles} \\
   \mbox{Initial parameters: large tan$\beta$ case ($\lambda = \xi$) }\ \ \
\ \

 (1/$\alpha_G, \, M_G, \, \epsilon_3$) = ($24.52, \, 3.05 \cdot 10^{16}$
GeV,$\,
-4.08$\%), \makebox[1.8em]{ }\\
 ($\lambda, \,$r$, \, \sigma, \, \epsilon, \, \rho, \, \epsilon^\prime$) =
($ 0.79, \,
12.4, \, 0.84, \, 0.011, \,  0.043,\,  0.0031$),\\
($\Phi_\sigma, \, \Phi_\epsilon, \, \Phi_\rho$) =  ($0.73, \, -1.21, \,
3.72$)rad,
\makebox[6.6em]{ }\\
($m_0, \, M_{1/2}, \, A_0, \, \mu(M_Z)$) = ($1000,\, 300, \, -1437, \,
110$) GeV,\\
($(m_{H_d}/m_0)^2, \, (m_{H_u}/m_0)^2, \, $tan$\beta$) = ($2.22,\, 1.65, \,
53.7$)
}
\label{t:fit4nu}
$$
\begin{array}{|l|c|l|}
\hline
{\rm Observable}  &{\rm Data}(\sigma) & Theory  \\
\mbox{ }   & {\rm (masses} & {\rm in\  \ GeV) }  \\
\hline
\;\;\;M_Z            &  91.187 \ (0.091)  &  91.17          \\
\;\;\;M_W             &  80.388 \ (0.080)    &  80.40       \\
\;\;\;G_{\mu}\cdot 10^5   &  1.1664 \ (0.0012) &  1.166     \\
\;\;\;\alpha_{EM}^{-1} &  137.04 \ (0.14)  &  137.0         \\
\;\;\;\alpha_s(M_Z)    &  0.1190 \ (0.003)   &  0.1174       \\
\;\;\;\rho_{new}\cdot 10^3  & -1.20 \ (1.3) & +0.320   \\
\hline
\;\;\;M_t              &  173.8 \ (5.0)   &  175.0       \\
\;\;\;m_b(M_b)          &    4.260 \ (0.11)  &    4.328                  \\
\;\;\;M_b - M_c        &    3.400 \ (0.2)   &    3.421                 \\
\;\;\;m_s              &  0.180 \ (0.050)   &  0.148          \\
\;\;\;m_d/m_s          &  0.050 \ (0.015)   &  0.0589        \\
\;\;\;Q^{-2}           &  0.00203 \ (0.00020)  &  0.00201                \\
\;\;\;M_{\tau}         &  1.777 \ (0.0018)   &  1.776         \\
\;\;\;M_{\mu}          & 0.10566 \ (0.00011)   & .1057           \\
\;\;\;M_e \cdot 10^3      &  0.5110 \ (0.00051) &  0.5110  \\
 \;\;\;V_{us}         &  0.2205 \ (0.0026)      &  0.2205        \\
\;\;\;V_{cb}         & 0.03920 \ (0.0030)      &  0.0403           \\
\;\;\;V_{ub}/V_{cb}    &  0.0800 \ (0.02)    &  0.0691                 \\
\;\;\;\hat B_K          &  0.860 \ (0.08)    &  0.8703           \\
\hline
{B(b\!\rightarrow\! s \gamma)\!\cdot\!10^{4}}  &  3.000 \ (0.47) &  2.995  \\
\hline
  \multicolumn{2}{|l}{{\rm TOTAL}\;\;\;\; \chi^2}  3.39
            & \\
\hline
\end{array}
$$
\end{table}

We have performed a global $\chi^2$ analysis, incorporating two (one) loop
renormalization group[RG] running of dimensionless (dimensionful)
parameters from
$M_G$ to $M_Z$ in the MSSM,  one loop radiative threshold corrections at $M_Z$,
and 3 loop QCD (1 loop QED) RG running below $M_Z$.  Electroweak symmetry
breaking is
obtained self-consistently from the effective potential at one loop, with
all one loop
threshold corrections included. This analysis is performed using the code
of Blazek
et.al.~\cite{blazek}.~\footnote{We assume universal scalar mass $m_0$ for
squarks and
sleptons at $M_G$.  We have not considered the flavor violating effects of U(2)
breaking scalar masses in this paper.}  In this paper, we just
present the results for one set of soft SUSY breaking parameters $m_0, \;
M_{1/2}$
with all other parameters varied to obtain the best fit solution.  In table
\ref{t:fit4nu} we give the 20 observables which enter the $\chi^2$
function, their
experimental values and the uncertainty $\sigma$ (in parentheses).   In most
cases
$\sigma$ is determined by the 1 standard deviation experimental uncertainty,
however in some cases the theoretical  uncertainty ($\sim$ 0.1\%) inherent in
our renormalization group running and one loop threshold corrections
dominates.

For large tan$\beta$ there are 6 real Yukawa parameters and 3 complex phases.
We take the complex phases to be $\Phi_\rho, \; \Phi_\epsilon$ and
$\Phi_\sigma$.  With 13 fermion mass observables (charged fermion masses and
mixing angles [$\hat{B}_K$ replacing $\epsilon_K$ as a ``measure of CP
violation"]) we
have 4 predictions.  For low tan$\beta$ , $\lambda \neq \xi$, we have one less
prediction.  From table \ref{t:fit4nu}  it is clear that this theory fits
the low energy data quite well.~\footnote{In
a future paper we intend to  explore the dependence of the fits on the SUSY
breaking
parameters and also U(2) flavor violating effects. Note also the strange
quark mass
$m_s(1 \rm GeV) \sim 150 \ \rm MeV$ is small, consistent with recent lattice
results.}    Note, fits with $\lambda \gg \xi$ and small $\tan\beta$ fit
just as
well.

Finally, the squark, slepton, Higgs and gaugino spectrum of our theory is
consistent with all available data.  The lightest chargino and neutralino are
higgsino-like with the masses close to their respective experimental
limits. As an
example of the additional predictions of this theory consider the CP
violating mixing
angles which may soon be observed at B factories.   For the selected fit we
find
\begin{eqnarray}
(\sin 2\alpha, \, \sin 2\beta, \, \sin \gamma) = & (0.74, \, 0.54, \,
0.99)&
\end{eqnarray}  or equivalently the Wolfenstein parameters
\begin{eqnarray}
(\rho, \, \eta )    =      &( -0.04, \,      0.31)    &.
\end{eqnarray}

\section{Neutrino Masses and Mixing Angles}

The parameters in the  Dirac Yukawa matrix for neutrinos (eqn.
\ref{eq:yukawa}) mixing
$\nu - \bar \nu$ are now fixed.  Of course, neutrino masses are much too
large and we need to invoke the GRSY~\cite{grsy} see-saw mechanism.

Since the {\bf 16} of SO(10) contains the  ``right-handed" neutrinos
$\bar \nu$, one possibility is to obtain  $\bar \nu - \bar \nu$ Majorana masses
via higher dimension operators of the form~\footnote{This possibility
has been considered in the paper by Carone and Hall~\cite{u2neutrino}.}
\begin{eqnarray}
{1 \over M} \ \overline{16} \ 16_3 \ \overline{16} \ 16_3   , \\
{1 \over M^2} \ \overline{16} \ 16_3 \ \overline{16} \ 16_a \ \phi^{a}  ,
\nonumber\\ {1 \over M^2} \ \overline{16} \ 16_a \ \overline{16} \ 16_b \
S^{a \, b}
. \nonumber
 \end{eqnarray}

The second possibility, which we follow, is to introduce SO(10) singlet
fields $N$ and obtain effective mass terms $\bar \nu - N$ and $N - N$
using only dimension four operators in the superspace potential.  To do this,
we add three new SO(10) singlets
\{$N_a,\; a = 1,2;\;\; N_3$\}  with U(1) charges \{  $-1/2$,\  +1/2 \}.
These then contribute to the superspace potential
\begin{eqnarray}
W \supset &  \overline{16} \; (N_a \; \chi^a \;\; + \;\; N_3 \; 16_3) &
\nonumber \\
 + &{1 \over 2} \ N_a \; N_b \; S^{a \; b} \;\; + \;\; N_a \; N_3 \; \phi^a
\end{eqnarray}
where the field $\overline{16}$ with U(1) charge  $-1/2$ is assumed to get a
vev in the ``right-handed" neutrino direction. Note, this vev is
also needed to break the rank of SO(10).

Finally we allow for the possibility of adding a U(2) doublet of SO(10)
singlets
$\bar N^a$ or a U(2) singlet $\bar N^3$.  They enter the superspace
potential as
follows --
\begin{eqnarray}
W \supset &  \mu' \; N_a \; \bar N^a \;\; + \;\;\mu_3 \;  N_3 \bar N^3 &
\label{eq:mu'}
\end{eqnarray}
The dimensionful parameters $\mu', \; \mu_3$ are assumed to be of order the
weak
scale.  The notation is suggestive of the similarity between these terms and
the $\mu$ term in the Higgs sector. In both cases, we are adding
supersymmetric mass
terms and in both cases, we need some mechanism to keep these dimensionful
parameters
small compared to the Planck scale.

We define the 3 $\times $ 3 matrix
\begin{eqnarray}
\tilde \mu = & \left( \begin{array}{ccc}  \mu' & 0 & 0 \\
                               0 & \mu' & 0 \\
                                 0  &  0  & \mu_3 \end{array}\right) &
\end{eqnarray}
The matrix $\tilde \mu$ determines the number of {\em coupled} sterile
neutrinos, i.e.
there are 4 cases labeled by the number of neutrinos ($N_\nu = 3, 4, 5, 6$):
\begin{itemize}
\item ($N_\nu = 3$) \hspace{.1cm} 3 active \hspace{.3cm} ($\mu' = \mu_3 =
0$);
\item ($N_\nu = 4$) \hspace{.1cm} 3 active + 1 sterile

\hspace{3cm} ($\mu' = 0;\; \mu_3 \neq 0$);
\item ($N_\nu = 5$) \hspace{.1cm} 3 active + 2 sterile

\hspace{3cm} ($\mu' \neq 0;\; \mu_3 = 0$);
\item ($N_\nu = 6$) \hspace{.1cm} 3 active + 3 sterile

\hspace{3cm} ($\mu' \neq 0;\; \mu_3 \neq 0$);
\end{itemize}
In this letter we consider the cases  $N_\nu = 3$ and 4~\cite{brt}.

The generalized neutrino mass matrix is then given by~\footnote{This
is similar to the double see-saw mechanism suggested by Mohapatra and
Valle~\cite{mohapatra}.}
\begin{eqnarray}
& ( \begin{array}{cccc}\; \nu & \;\; \bar N & \;\; \bar \nu & \;\;  N
\end{array})  &
\nonumber\\  &  \left( \begin{array}{cccc}  0 & 0 & m & 0  \\
                     0 & 0 & 0 & \tilde \mu^T \\
                     m^T & 0 & 0 & V \\
                     0 & \tilde \mu & V^T & M_N  \end{array} \right) &
\end{eqnarray}
where  \begin{eqnarray} m = & Y_{\nu}\; \langle H_u^0 \rangle
&= \; Y_{\nu}\; {v \over\sqrt{2}}\; \sin\beta \end{eqnarray} and
\begin{eqnarray}
V = & \left( \begin{array}{ccc}  0 & \epsilon' V_{16} & 0 \\
                                - \epsilon' V_{16} & 3 \epsilon V_{16} & 0 \\
                                 0  & r \, \epsilon \, (1 - \sigma) \,
T_{\bar \nu} V_{16}  &
V'_{16}
\end{array}\right) &
\\
  M_N = & \left( \begin{array}{ccc}  0 & 0 & 0 \\
                                0 & S & \phi \\
                                 0  & \phi  &  0 \end{array}\right) &
\nonumber
\end{eqnarray}
$V_{16},\; V'_{16}$ are proportional to the vev of $\overline{16}$
(with different
implicit Yukawa couplings) and $S, \; \phi$ are up to couplings the vevs of $
S^{22}, \; \phi^2$, respectively.

Since both $V$ and $M_N$ are of order the GUT scale, the states $\bar \nu,
\; N$
may be integrated out of the effective low energy theory.  In this case, the
effective neutrino mass matrix is given (at $M_G$) by~\footnote{In fact, at the
GUT scale $M_G$ we define an effective dimension 5 supersymmetric neutrino mass
operator where the Higgs vev is replaced by the Higgs doublet  H$_u$ coupled
to the entire lepton doublet.  This effective operator is then renormalized
using one-loop
renormalization group equations to $M_Z$.  It is only then that $H_u$ is
replaced by its
vev.} (the matrix is written in the ($\nu, \ \bar N$) {\em flavor} basis
where  charged lepton masses are diagonal)
\begin{eqnarray}
 m_\nu =  & \hspace{2in}  &
\end{eqnarray}
\begin{eqnarray}
 \tilde U_e^T \; \left( \begin{array}{cc}
m\;(V^T)^{-1}\;M_N\; V^{-1}\; m^T &  - m \;(V^T)^{-1}\; \tilde \mu\\
                    - {\tilde \mu}^T \; V^{-1}\; m^T & 0  \end{array}
\right) \; \tilde
U_e \nonumber
\end{eqnarray}
with
\begin{eqnarray} \tilde U_e = & \left(\begin{array}{cc} U_e & 0 \\
                                               0 & 1 \end{array}\right) & \\
e_0 =  U_e \; e \;\; ; &  \nu_0 = U_e \; \nu  &  \nonumber
\end{eqnarray}
 $U_e$ is the $3\times3$ unitary matrix for left-handed leptons needed to
diagonalize $Y_e$ (eqn. \ref{eq:yukawa}) and $e_0,\; \nu_0 \; (e, \; \nu)$
represent the
three families of left-handed leptons in the charged-weak ( -mass)
eigenstate basis.

The neutrino mass matrix is diagonalized by a unitary matrix $U =
U_{\alpha\, i}$;
\begin{eqnarray}
m^{diag}_{\nu} = & U^T \; m_{\nu} \; U &
\end{eqnarray}
where $\alpha= \{\nu_e ,\; \nu_\mu ,\; \nu_\tau ,\; \nu_{s_1}, \;
\nu_{s_2}, \; \nu_{s_3} \}$ is the flavor index and
$i = \{ 1, \cdots, 6\}$ is the neutrino mass eigenstate index.
$U_{\alpha\, i}$   is
observable in neutrino oscillation experiments.    In particular,  the
probability
for the flavor state $\nu_\alpha$ with energy $E$ to oscillate into
$\nu_\beta$ after
travelling a distance $L$ is given by
\begin{eqnarray}
&&P(\nu_\alpha \rightarrow \nu_\beta)  = \\&& \delta_{\alpha \beta}
- 4\sum_{k\; <\, j} U_{\alpha \, k} \ U^*_{\beta \, k} \
U^*_{\alpha \, j} \ U_{\beta \, j} \ \sin^2\Delta_{j\, k}  \nonumber
\end{eqnarray}
where $\Delta_{j\,k} =  {\delta m^2_{j k} \ L \over 4 E}$ and
$\delta m^2_{j k} = m^2_j - m^2_k$.

In general, neutrino masses and mixing angles have many new parameters so
that one
might expect to have little predictability.   However, as we shall now see, the
U(2)$\times$U(1) flavor symmetry of the theory provides a powerful
constraint on
the form of the neutrino mass matrix.  In particular, the matrix has many
zeros and few
arbitrary parameters.  Before discussing the four neutrino case, we show why 3
neutrinos cannot work without changing the model.

\subsection{Three neutrinos}

Consider first $m_{\nu}$ for three active neutrinos.  We find (at $M_G$) in
the ($\nu_e,\; \nu_\mu,\; \nu_\tau$) basis
\begin{eqnarray}
m_{\nu} =  m' \; U_e^T \; \left(\begin{array}{ccc}  0 & 0 & 0 \\
                                       0 & b & 1  \\
                                       0 & 1 & 0  \end{array} \right) \; U_e
\end{eqnarray}
with
\begin{eqnarray}  m' &=& \frac{\lambda^2 v^2 \sin^2\beta~ \omega \phi}
{2 V_{16} V'_{16}} \approx {m_t^2 \ \omega \ \phi  \over
V_{16} \ V'_{16}}  \label{eq:3nu} \\
b &= &  \omega \ {S \ V'_{16} \over \phi \ V_{16}} + 2 \ \sigma \ r \ \epsilon
 \nonumber
   \end{eqnarray}
where in the approximation for $m'$ we use
\begin{eqnarray}   m_t (\equiv m_{top}) \approx \lambda \ {v \over \sqrt{2}}
\ \sin\beta ,
\end{eqnarray}  valid at the weak scale.

$m_{\nu}$ is given in terms of two independent parameters \{ $m', \; b$ \}.
Note, this theory in principle solves two problems associated with neutrino
masses.   It naturally has small mixing between $\nu_e - \nu_{\mu}$ since
 the mixing angle comes purely from diagonalizing the charged
lepton mass matrix which, like quarks, has small mixing angles.   While, for
$b \leq  1$, $\nu_{\mu} - \nu_{\tau}$ mixing is large without fine tuning.
Also note, in this theory one neutrino (predominantly $\nu_e$) is
massless.

Unfortunately this theory cannot simultaneously fit both solar and atmospheric
neutrino data.   This problem can be solved at the expense of
 adding a new flavor symmetry breaking vev~\footnote{This additional vev
 was necessary in the analysis of Carone and Hall~\cite{u2neutrino}.}
 \begin{eqnarray}  \langle \phi^1 \rangle = \kappa \langle \phi^2\rangle  .
  \label{eq:kappa} \end{eqnarray}
We discuss this three neutrino
 solution in a future paper~\cite{brt}.  With $\kappa \neq 0$ the massless
eigenvalue
in the neutrino mass matrix is  now lifted.  This allows us to obtain a
small mass
difference between the  first and second mass eigenvalues which was
unattainable
before in the large  mixing limit for $\nu_\mu - \nu_\tau$.  Hence a good
fit to both
solar and  atmospheric neutrino data can now be found for $\kappa \leq
0.2$.  In
addition, note that this small value of $\kappa$ moderately improves the global
fits to
charged fermion masses and mixing angles\cite{brt}.

In the next section we discuss a four neutrino solution to both solar and
atmospheric
neutrino oscillations in the theory with $\kappa = 0$.

 \subsection{Neutrino oscillations   [ 3 active + 1 sterile ]}

 In the four neutrino case the mass matrix (at $M_G$) is given
by~\footnote{This
expression defines the effective dimension 5 neutrino mass operator at
$M_G$ which
is then renormalized to $M_Z$ in order to make contact with data.}

 \begin{eqnarray} m' \  \left[  \begin{array}{cccc}
     U_e^T \,   \left( \begin{array}{ccc}   0 & 0 & 0 \\
                                       0 & b  & 1  \\
                                       0 & 1 & 0  \end{array} \right) \, U_e &
- U_e^T \left( \begin{array}{c} 0 \\ u \, c \\ c \end{array} \right) \\
- \left( \begin{array}{ccc} 0 & u \, c & c \end{array} \right) \, U_e  & 0
\end{array}
\right]
\end{eqnarray}
where  $m'$ and $b$ are given in eqn. \ref{eq:3nu}  and
\begin{eqnarray}  u &=& \sigma \, r \, \epsilon \label{eq:c} \\
   c &= & {\sqrt{2} \  \mu_3 \, V_{16} \over \omega \,  \lambda \ v \
	\sin\beta \ \phi}
 \approx  { \mu_3 \, V_{16} \over \omega \,  m_t \ \phi}  \nonumber
\end{eqnarray}

In the analysis of neutrino masses and mixing angles we use the fits for
charged fermion
masses as input.  Thus the parameter $u$ is fixed.  We then look for the
best fit to solar
and atmospheric neutrino oscillations.   For this we use the latest
Super-Kamiokande data for atmospheric neutrino oscillations~\cite{atmos} and
the best
fits to solar neutrino  data including the possibility of ``just so" vacuum
oscillations or both large and small angle MSW oscillations~\cite{solar}.
Our best fit
is found in tables \ref{t:4numass2} and \ref{t:4nuangles}.  It is obtained
in the
following way.

For atmospheric neutrino oscillations we have evaluated the probabilities
($P(\nu_\mu \rightarrow \nu_\mu)$,   $P(\nu_\mu \rightarrow \nu_x) \ {\rm
with} \ x = \{ e, \ \tau, \ s \}$)  as a function of ${\rm x} \equiv
\rm{Log}[(L/km)/(E/GeV)]$.  In order to smooth out the oscillations we have 
averaged the result over a bin size, $\Delta$x = 0.5.   In  fig. 2a we have 
compared the results of our model with a 2 neutrino oscillation model.  We 
see that our result is in good agreement with the
values of $\delta m^2_{atm}$ and $\sin^2 2 \theta_{atm}$ as given.

An approximate formula for the effective atmospheric mixing angle is defined by
\begin{eqnarray}
P(\nu_\mu \rightarrow \nu_\mu) \equiv 1- `\sin^2 2 \theta_{atm}` \sin^2
({`\delta m^2_{atm}` \ \rm L \over 4 \ E})  \end{eqnarray}
 with
\begin{eqnarray} `\sin^2 2 \theta_{atm}` &\approx &  4 \ [ \ \| U_{\mu 4}\|^2
( 1 - \| U_{\mu 4} \|^2) \\ 
&+&\| U_{\mu 3} \|^2 ( 1 - \| U_{\mu 3} \|^2 - \| U_{\mu 4} \|^2 )\ ]  \nonumber
\end{eqnarray}
using the approximate relation
\begin{eqnarray}  `\delta m^2_{atm}` = \delta m^2_{43} \approx \delta m^2_{42}
\approx \delta m^2_{41}
\approx \delta m^2_{32} \approx \delta m^2_{31}   .
\end{eqnarray}
Note, $`\sin^2 2 \theta_{atm}` $ may be greater than one.  This is
consistent with the definition above and also with Super-Kamiokande data 
where the best fit occurs for
$\sin^2 2\theta_{atm} = 1.05$.  We obtain a good fit to the data.

In fig. 2b we see however that although the atmospheric neutrino deficit 
is predominantly due to the maximal mixing
between $\nu_\mu - \nu_\tau$, there is nevertheless a significant ($\sim$ 10\%
effect)  oscillation of $\nu_\mu - \nu_s$.  This effect may be observable at
Super-Kamiokande.  It would appear as a deficit of neutrinos in the ratio of
experimental to theoretical muon (single ring events) plus tau
(multi-ring events) as a function of $L/E$.

The oscillations $\nu_\mu \rightarrow \nu_\tau$ or $\nu_s$ may also be
visible at long baseline neutrino experiments.  For example at K2K~\cite{k2k},
  the mean neutrino energy $E = 1.4 $GeV and distance $L = 250$ km
corresponds to a value of x = 2.3 in fig. 2b and hence  $P(\nu_\mu \rightarrow
\nu_\tau) \sim .4$ and $P(\nu_\mu \rightarrow \nu_s) \sim .1$.  At 
Minos~\cite{minos} low energy beams with hybrid emulsion detectors
 are also being considered. 
These experiments can first test the hypothesis of muon neutrino oscillations
by looking for muon neutrino disappearance (for x = 2.3 we have
$P(\nu_\mu \rightarrow \nu_\mu) \sim .5$).  Verifying oscillations into sterile 
neutrinos is however much more difficult.
For example at K2K, if only quasi-elastic muon neutrino interactions (single 
ring events at SuperK) are used, then this cannot be tested.  
Minos, on the other hand, may be able to verify the oscillations into sterile 
neutrinos by using the ratio of neutral current to charged current
measurements~\cite{minos} (the so-called T test). 
 
 \protect
\begin{table}
\caption[3]{
{\bf Fit to atmospheric and solar neutrino \\ oscillations} \\
   \mbox{Initial parameters: ( 4 neutrinos \  $w$/ large tan$\beta$  ) }\ \
\ \

$m' = 7.11 \cdot 10^{-2}$ eV , \ $b$ = $-0.521$, \ $c$ = 0.278,\ $\Phi_b$ =
3.40rad
}
\label{t:4numass2}
$$
\begin{array}{|c|c|}
\hline
{\rm Observable} &{\rm Computed \;\; value} \\
\hline
\delta m^2_{atm}            &  3.2 \cdot 10^{-3} \ \rm eV^2          \\
\sin^2 2\theta_{atm}            &  1.08        \\
 \delta m^2_{sol}   &  4.2\cdot 10^{-6}  \ \rm eV^2  \\
\sin^2 2\theta_{sol} &  3.0\cdot 10^{-3}         \\
\hline
\end{array}$$
\end{table}

\protect
\begin{table}
\caption[3]{
{\bf Neutrino Masses and Mixings} \hspace{1.1cm} \\

\mbox{Mass eigenvalues [eV]: \ \  0.0, \ 0.002, \ 0.04, \ 0.07 \hspace{1cm}} \\
\mbox{Magnitude of neutrino mixing matrix  U$_{\alpha i}$ \hspace{1.7cm}}\\
\mbox{ $i = 1, \cdots, 4$ -- labels mass eigenstates. \hspace{1.5cm}} \\
\mbox{ $\alpha = \{ e, \ \mu, \ \tau, \ s \}$ labels flavor eigenstates.}
}
\label{t:4nuangles}
$$
\left[ \begin{array}{cccc}
0.998                   &  0.0204      & 0.0392   & 0.0529  \\
0.0689                  &    0.291     & 0.567    & 0.767  \\
0.317\cdot 10^{-3}      &  0.145       & 0.771    & 0.620  \\
0.284\cdot 10^{-3}      &   0.946      &  0.287     &  0.154 \\
\end{array} \right]$$
\end{table}

\renewcommand{\thefigure}{2 \alph{figure}}\setcounter{figure}{0}
\begin{figure}
	\centerline{ \psfig{file=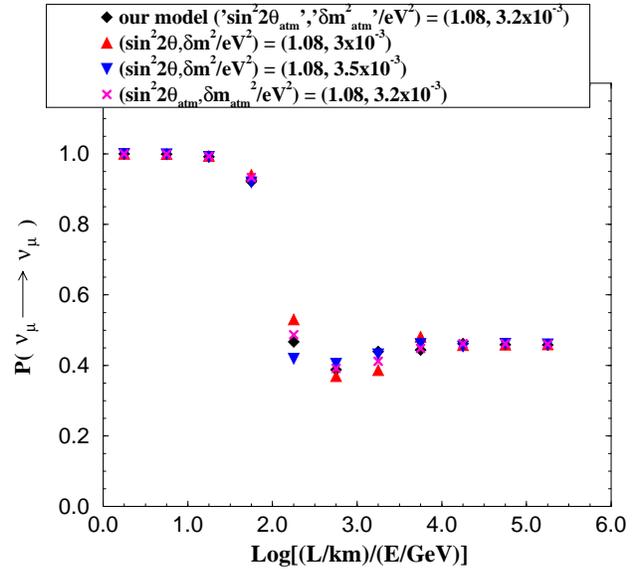,width=9cm,angle=-90}}
\caption{Probability $P(\nu_\mu \longrightarrow \nu_\mu)$ for atmospheric
neutrinos. For this analysis, we neglect the matter effects.}
\end{figure}
\begin{figure}
	\centerline{ \psfig{file=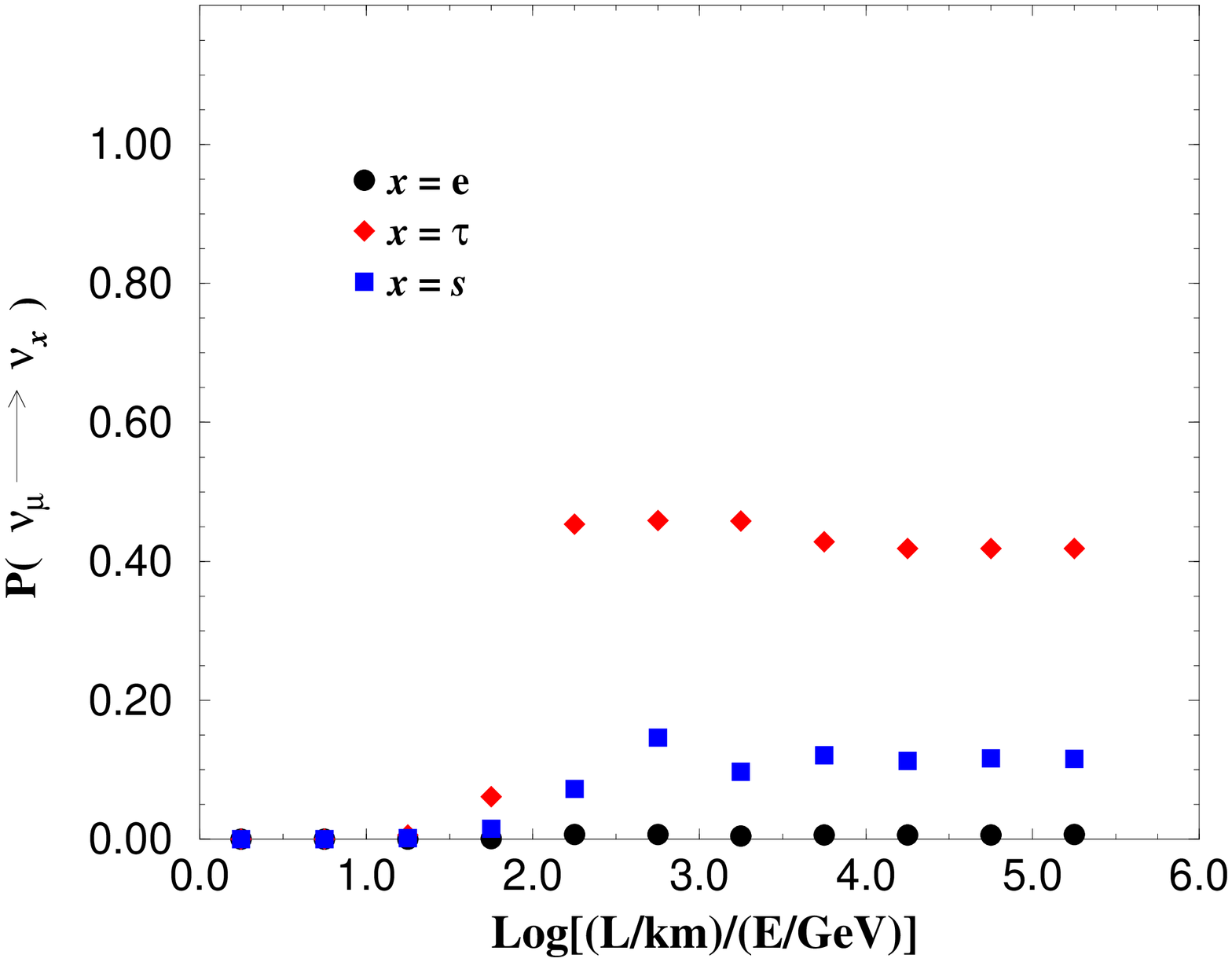,width=9cm,angle=-90}}
\caption{Probabilities $P(\nu_\mu \longrightarrow \nu_x)$ ($x=e$, $\tau$ and
$s$) for atmospheric neutrinos}
\end{figure}

For solar neutrinos we plot,  in figs. 3(a,b), the probabilities  ($P(\nu_e
\rightarrow \nu_e)$,  $P(\nu_e \rightarrow \nu_x) \ {\rm with}
\ x = \{ \mu, \ \tau, \ s \}$) for neutrinos produced at the center of the
sun to propagate
to the surface (and then without change to earth), as a function of the
neutrino energy  E$_\nu$ (MeV).~\footnote{For this calculation
we assume that electron ($n_e$) and neutron ($n_n$) number densities
at a distance $r$ from the center of the sun are given by
$(n_e,~n_n)=(4.6,2.2)\times 10^{11} \exp(-10.5 \frac{r}{R})$ eV$^3$
where $R$ is a solar radius. We also use an analytic approximation necessary to
account for both large and small oscillation scales. For the details, see
the forthcoming paper\cite{brt}.}  We compare our model to a 2 neutrino
oscillation model
with the given parameters.   We see that the solar neutrino deficit is
predominantly due to the small mixing angle MSW solution for $\nu_e -
\nu_s$ oscillations.  The results are summarized in tables
\ref{t:4numass2} and \ref{t:4nuangles}.

A naive definition of the effective  solar mixing angle is given by
\begin{eqnarray} \sin^2 2 \theta_{12} \equiv  4 \ \| U_{e 1} \|^2 \ \|
U_{e 2} \|^2 .
\end{eqnarray}
In fig. 3a we see that the naive definition of $\sin^2 2 \theta_{12} $
underestimates the value of the effective 2 neutrino
mixing angle.   Thus we see that our model reproduces the neutrino results for
$\delta m^2_{sol} = \delta m^2_{12} = 4.2 \times 10^{-6}$ eV$^2$ but instead is
equivalent to a 2 neutrino mixing angle
$\sin^2 2 \theta_{sol} =  3 \times 10^{-3}$
instead of $\sin^2 2 \theta_{12} = 1.6 \times 10^{-3}$.
Our result is consistent with the fits of Bahcall et al.~\cite{solar}.

In addition, whereas the oscillation
$\nu_e - \nu_s$ dominates we see in fig 3b that there is a sigificant
($\sim$ 8\% effect) for  $\nu_e - \nu_\mu$.  This result may be observable at
SNO~\cite{sno}  with threshold $E \ge 5$ MeV for which $P(\nu_e \rightarrow
\nu_\mu) \sim .05$.

\renewcommand{\thefigure}{3 \alph{figure}}\setcounter{figure}{0}
\begin{figure}
	\centerline{ \psfig{file=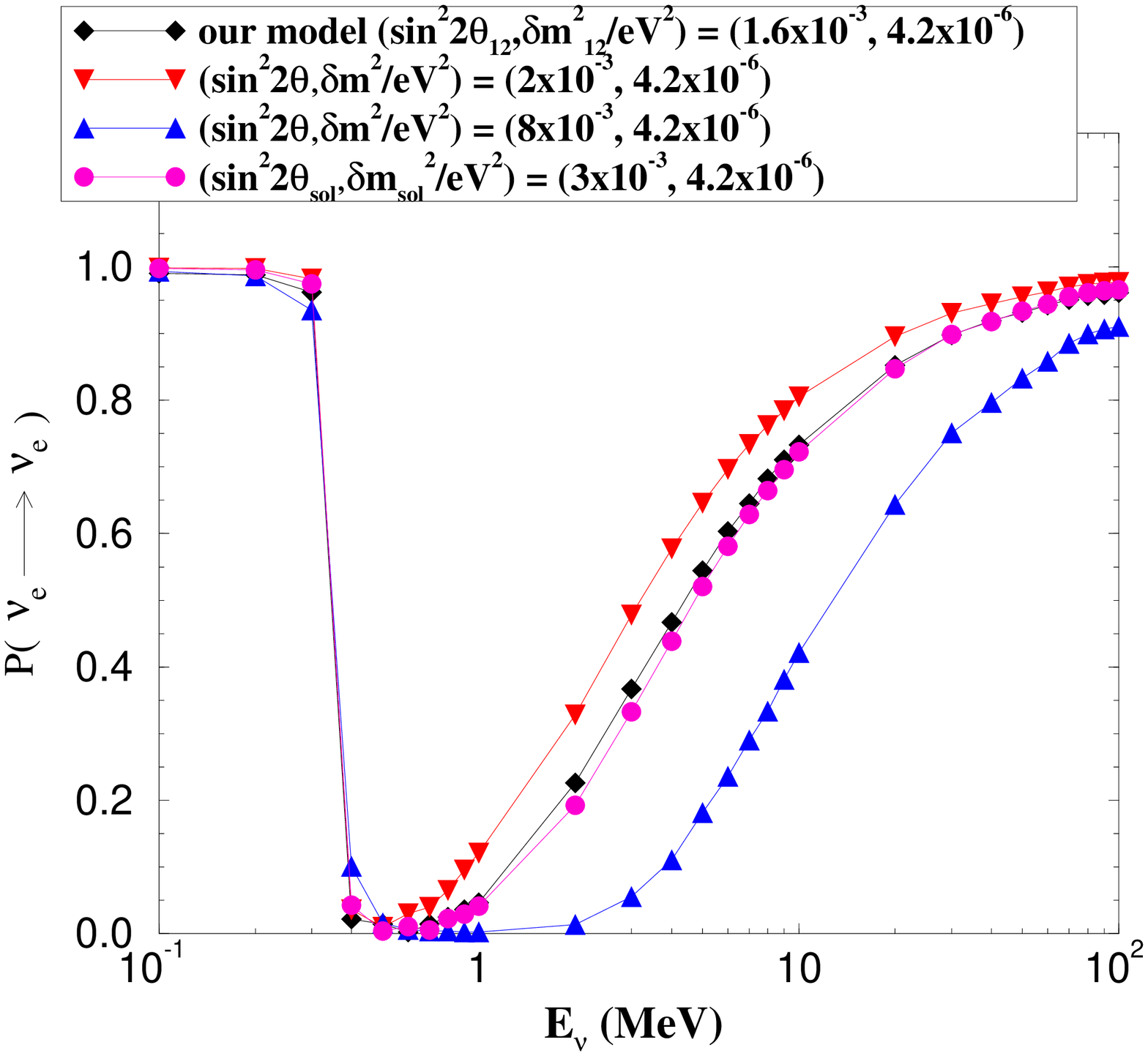,width=9cm,angle=-90}}
\caption{Probability $P(\nu_e \longrightarrow \nu_e)$ for solar neutrinos}
\end{figure}
\begin{figure}
	\centerline{ \psfig{file=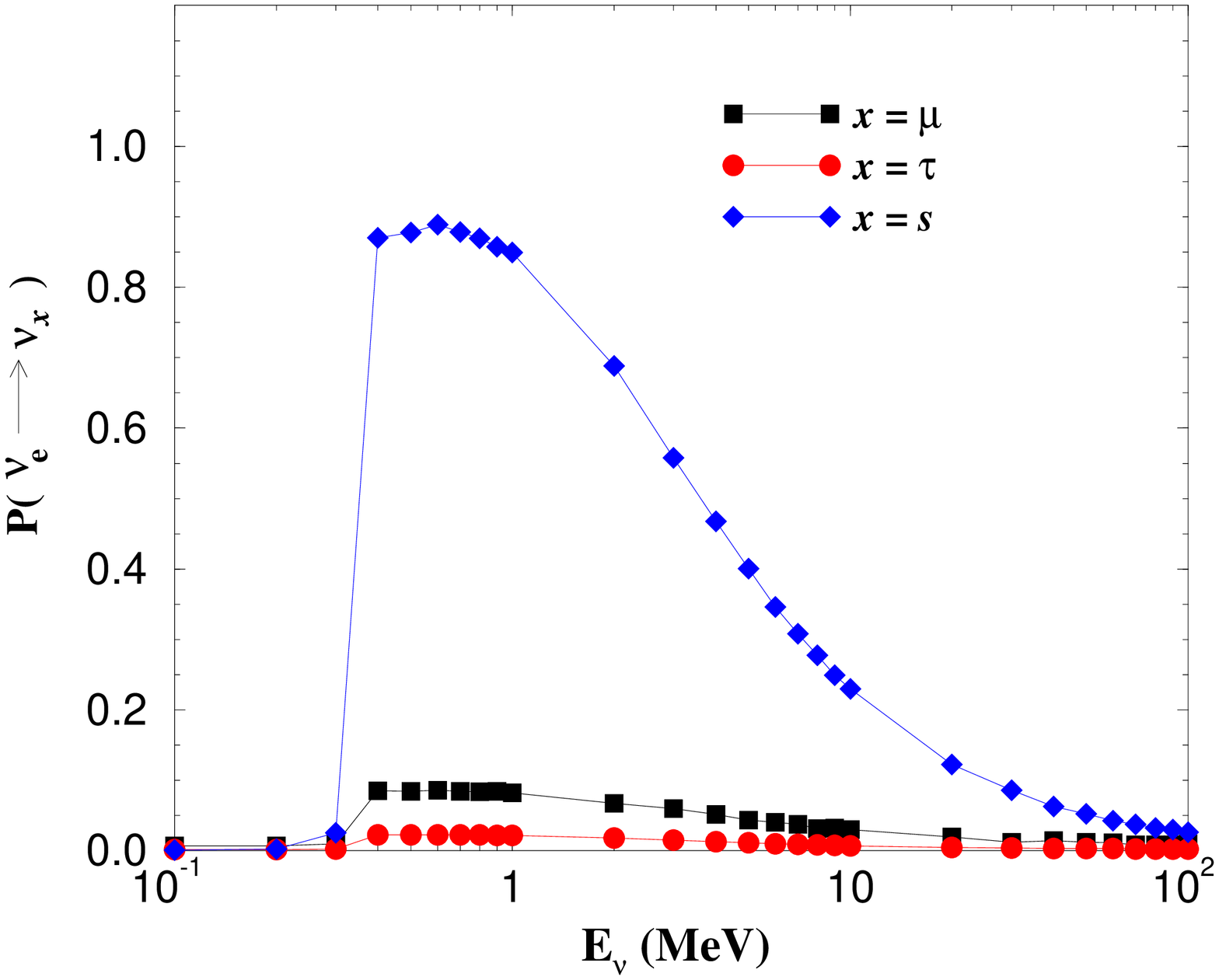,width=9cm,angle=-90}}
\caption{Probabilities $P(\nu_e \longrightarrow \nu_x)$ ($x=\mu$, $\tau$ and
$s$)
for solar neutrinos}
\end{figure}

{\em We note that, even though we have four neutrinos, we are {\bf not} able to
simultaneously fit atmospheric, solar and LSND data, i.e. it is not
possible to get
$``\delta m^2_{\nu_e - \nu_\mu}"$ large enough to be consistent with LSND.
We have
also checked that introducing the new parameter $\kappa$ (eqn.
\ref{eq:kappa}) does
not help.}

Finally let's discuss whether the parameters necessary for the fit make sense.
We have three arbitrary parameters.  We have taken $b$ complex,
while any phases for $m'$ and $c$ are unobservable.
A large mixing angle for $\nu_\mu - \nu_\tau$ oscillations is
obtained with $|b| \sim 0.5$.  This does not require any fine tuning; it is
consistent with ${S \ V'_{16} \over \phi \ V_{16}} \sim 1$ which is perfectly
natural (see eqn. \ref{eq:3nu}).
The parameter $c [\rm{eqn.} \ref{eq:c}] \approx 0.28 \approx
{ \mu_3 \, V_{16} \over \omega \,
m_t \ \phi}$ implies $ \mu_3 \approx 0.41 \ {\phi \over V_{16}} \ m_t$.
Thus in order to have a light sterile neutrino we need the parameter $\mu_3
\sim 110$ GeV for $\phi \sim V_{16}$.   Considering that the standard $\mu$
parameter
(see the parameter list in the captions to table \ref{t:fit4nu}) with value
$\mu = 110$ GeV and $\mu_3$ [eqn. \ref{eq:mu'}] may have similar origins,
both generated once SUSY
is spontaneously broken, we feel that it is natural to have a light sterile
neutrino.  Lastly consider the overall scale of
symmetry breaking, i.e. the see-saw scale.  We have $ m' =  7 \times
10^{-2} \rm eV [table ~\ref{t:4numass2}] \approx
{m_t^2 \ \omega \ \phi  \over  V_{16} \ V'_{16}} $.  Thus we find
$ {V_{16} \ V'_{16} \over \phi} \sim  {m_t^2 \ \omega \over m'} \sim
1.6 \times 10^{15}$ GeV.  This is perfectly reasonable for
$\langle \overline{16} \rangle \sim \langle \phi^2 \rangle \sim M_G$
once the implicit Yukawa couplings are taken into account.

\section{Conclusion}

We have presented the results of a predictive
SO(10)$\times$U(2)$\times$U(1) model of
fermion masses.   We fit charged fermion masses and mixing angles as well as
neutrino masses and mixing.  The model ``naturally" gives small mixing
angles  for
charged fermions and for $\nu_e \rightarrow \nu_{sterile}$ oscillations
(small angle
MSW solution to solar neutrino problem) and large mixing angle for $\nu_\mu
\rightarrow
\nu_\tau$ oscillations (atmospheric muon neutrino deficit).  The model
presented here
may be one of a large class of models which fit charged fermion masses.  The
most
important conclusion from our work is that predictive theories of charged
fermion
masses (including GUT and family symmetry) strongly constrain the neutrino
sector of
the theory.  These theories can thus be predictive in the neutrino sector
and neutrino
data will strongly constrain any predictive theory of fermion masses.

{\bf Acknowledgements}
Finally, S.R. and K.T. are partially supported by DOE grant DOE/ER/01545-761.

%

\end{document}